# Superoxide Reductase SOR from *Desulfoarculus baarsii*

By Vincent Nivière* and Murielle Lombard

Laboratoire de Chimie et Biochimie des Centres Redox Biologiques, DBMS-CEA/CNRS/Université Joseph Fourier, 17 Avenue des Martyrs, 38054 Grenoble, Cedex 9, France.

Running head : Superoxide Reductase from *D. baarsii*

* To whom correspondence should be addressed. Tel. : 33-4-76-88-91-09;  Fax : 33-4-76-88-91-24; E-mail : vniviere@cea.fr

# Introduction : Superoxide Reductase as a new enzymatic system involved in superoxide detoxication

Superoxide radical ($O_2^-$) is the univalent reduction product of molecular oxygen and belongs to the group of the so-called toxic oxygen derivatives. For years the only enzymatic system known to catalyze the elimination of superoxide was the superoxide dismutase (SOD) [1], which catalyzes dismutation of superoxide radical anions to hydrogen peroxide and molecular oxygen :

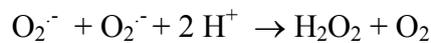
$$O_2^- + O_2^- + 2\,H^+ \rightarrow H_2O_2 + O_2$$

Very recently, a new concept in the field of the mechanisms of cellular defense against superoxide has emerged. [2, 3] It was discovered that elimination of $O_2^-$ could also occur by reduction, a reaction catalyzed by an enzyme thus named superoxide reductase (SOR) :

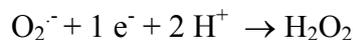
$$O_2^- + 1\,e^- + 2\,H^+ \rightarrow H_2O_2$$

Up to now, SOR has been mainly characterized from anaerobic microorganisms, sulfate-reducing bacteria [2], archaeon [3] and in a microaerophilic bacterium. [4, 5] In vivo, SOR was shown to be an efficient antioxidant protein from the observation that a *Desulfovibrio vulgaris Hildenborough* mutant strain lacking the *sor* gene became more oxygen-sensitive during transient exposure to microaerophilic conditions. [6] In addition, although SOR is not naturally present in *Escherichia coli*, it was demonstrated that expression of SORs from *Desulfoarculus baarsii* and *Treponema pallidum* in a *sodA sodB E. coli* mutant strain could totally replace the SOD enzymes to overcome a superoxide stress. [4, 5, 7]

Two classes of SOR have been described from now. Class I SORs are small metalloproteins found in anaerobic sulfate-reducing and microaerophilic bacteria, initially called desulfoferrodoxins (Dfx). There are homodimers of 2x14kDa, which have been extensively studied for their structural properties. [8, 9] The monomer is organized in two protein domains. [9] The N-terminal domain contains a mononuclear ferric iron, Center I, coordinated by four cysteines in a distorted rubredoxin-type center. In the SOR from *Treponema pallidum*, three of the four N-terminal cysteine residues involved in iron chelation are lacking and then Center I is missing. [4, 5] The C-terminal domain, which carries the active site of SOR [2], contains a different mononuclear iron center, Center II, consisting of an oxygen-stable ferrous iron with square-pyramidal coordination to four nitrogens from histidines as equatorial ligands and one sulfur from a cysteine as the axial ligand. [9] Its midpoint redox potential has been reported to be around +250 mV. [2, 5] The iron Center II reduces superoxide very efficiently, with a second order rate constant of about $10^9$ $M^{-1}$ $s^{-1}$. [2] It does not exhibit a significant SOD activity. [2] In addition, the active site of SOR is specific for $O_2^-$, since the reduced iron Center II is not oxidized by $O_2$ and only very slowly by $H_2O_2$. [2]

Class II SOR has been characterized from the anaerobic archaeon, *Pyrococcus furiosus*. [2, 10] The homotetrameric protein presents strong homologies to neelaredoxin (Nlr), a small protein containing a single mononuclear center, earlier characterized from sulfate-reducing bacteria. The amino acid sequence and the overall protein fold is similar to that of the iron Center II domain of Class I SORs and the structure of the unique mononuclear iron center is similar to that of the iron Center II. [10] The main difference is that Class II SOR do not contains the N-terminal domain that chelates the iron Center I in Class I SORs.

# SOR from *D. baarsii* : Class I SOR

*Overexpression and purification of the enzyme*

SOR from *Desulfoarculus baarsii* has been overexpressed in *Escherichia coli* using a vector placing its structural gene under the control of a *tac* promoter. [2] Induction was achieved by adding 1 mM IPTG at the beginning of the exponential phase, and best expression was obtained when the cells reach the stationary phase. SOR was overexpressed to about 10-15 % of the total soluble proteins from *E. coli*. No formation of inclusion bodies was noticed. Metallation of SOR was best achieved when the cells were grown on minimal media M9, complemented with 0.4 % glucose and 100 µM $FeCl_3$. In these conditions, SOR was fully metallated. Overexpression of SOR in Luria Bertani medium resulted in a protein partially demetallated.

Purification of SOR was followed by UV-visible spectrophotometry and SDS-PAGE electrophoresis. The ratio $A_{280nm}/A_{503nm}$, characteristic of the oxidized iron Center I of SOR, increases during purification until it reaches a factor of 7 for the purified protein. During purification, the iron Center II remains in a reduced state and does not contribute to the visible spectrum. It can be oxidized with a slight excess of $Fe(CN)_6$ and then exhibits an absorbance band centered at 644 nm (Figure 1). Purified SOR exhibits a value of $\varepsilon_{503nm}$ of 4,400 $M^{-1}$ $cm^{-1}$. Difference spectrum of the $Fe(CN)_6$ oxidized minus the as isolated SOR provides the absorption bands associated with the iron Center II with a $\varepsilon_{644nm}$ of 1,900 $M^{-1}$ $cm^{-1}$ (Figure 1). It should be noticed that SDS-PAGE analysis of the fully purified SOR exhibits several polypeptide bands. These artifacts could arise from partial oxidation of cysteine residues during electrophoresis migration. A single polypeptide band is obtained when the reduced thiols were omitted from the SDS-PAGE loading buffer.

Purification of the overexpressed SOR can be usually achieved with two chromatography steps. [2] In a typical experiment, 20 g of IPTG-induced *E. coli* cells are sonicated in 70 ml of 0.1 M Tris/HCl, pH 7.6 and centrifuged at 180,000xg for 90 min. Soluble extracts (400 mg) are first treated with 3% streptomycin sulfate, centrifuged at 40,000xg for 15 min and then precipitated with 75% ammonium sulfate. The pellet is dissolved in 0.1 M Tris/HCl, pH 7.6 and the solution is loaded onto a gel filtration ACA 54 column (360 ml) equilibrated with 10 mM Tris/HCl, pH 7.6 (buffer A). The low molecular weigh fractions with a $A_{280nm}/A_{503nm}$ ratio of about 20 are pooled (115 mg) and loaded into an anion exchange UNO-Q column (6 ml, Bio-Rad), equilibrated with buffer A. A linear gradient is applied (0-0.2 M NaCl in buffer A) for 60 ml. 30 mg of pure, fully metallated SOR is eluted with about 50 mM NaCl.

*Assay for SOR activity*

The reaction catalyzed by SOR can be described by the sum of two half reactions at its active site :

$$SOR_{red} + O_2^{\cdot -} + 2\,H^+ \rightarrow H_2O_2 + SOR_{ox} \quad (1)$$
$$SOR_{ox} + 1\bar{e} \rightarrow SOR_{red} \quad (2)$$
$$\overline{O_2^{\cdot -} + 1\,\bar{e} + 2\,H^+ \rightarrow H_2O_2 \quad (3)}$$

In the first half reaction (Eq. (1)), the reduced iron Center II of SOR transfers one electron to $O_2^{\cdot -}$ to produce $H_2O_2$ and the oxidized form of the iron Center II. This reaction is very rapid and occurs near the limit of diffusion of molecules in solution. In the second half reaction (Eq. (2)), the oxidized iron Center II is reduced by an electron donor to regenerate the active form of SOR. The physiological electron donors for SOR from *D. baarsii* are not identified yet, although in *E. coli*, it was found that the NADPH flavodoxin reductase (Fpr) can provide efficiently electrons to SOR (Manuscript in preparation).

Direct measurement of the global SOR activity (Eq. (3)) may be difficult for several reasons. The absolute requirement of $O_2$ in the assay to generate $O_2^{\cdot -}$ (usually generated by the xanthine-xanthine oxidase system) may strongly interfere with the electron donor. In addition, a possible use of non-autoxidable electron donors, like reduced cytochrome c, may not give an accurate measurement of a full SOR activity. As a matter of fact, the overall rate constant for the global reaction is rate limited by the electron transfer step to the oxidized form of SOR that would generally occurs 100 to 1000 time more slowly that the reduction of $O_2^{\cdot -}$ by SOR.

Then measurement of a global SOR activity will only provide kinetic informations of the reduction process of SOR by the electron donor rather than that of the reaction with superoxide.

All these remarks make difficult or almost impossible the measurement of the SOR activity in crude extracts. However, for a purified SOR preparation, it is possible to measure its ability to reduce superoxide by the following assay.

We have developed a SOR assay which allows to determine the rate constant of the first half reaction of SOR (Eq.(1)), corresponding to a stoechiometric reduction of superoxide by SOR.[2] We have used a methodology developed in the case of several dehydratases, such as aconitase and fumarase, which are known to react with $O_2^-$ very rapidly.[11] The general principle of this SOR assay is to follow spectrophotometrically the kinetic of oxidation of SOR by superoxide, generated by the xanthine-xanthine oxidase system. This is possible at 644 nm which is a characteristic absorption band of the oxidation of the active site of SOR (Figure 1). As shown in Figure 2, the kinetics are followed for several minutes, in the absence or in the presence of different amount of SOD, which compete with SOR for superoxide. Sufficient amount of SOD induces an inhibition of the kinetic of oxidation of SOR. As explained in detail elsewhere[2], in these experimental conditions, the velocity of oxidation of the active site of SOR by $O_2^-$ ($v_{ox}$) can be expressed as follow :

$$\frac{1}{v_{ox}} = \frac{1}{cte} + \frac{k_{SOD}}{cte\, k_{SOR}[SOR]}[SOD] \quad (4)$$

where $k_{SOR}$, $k_{SOD}$ are the second order rate constants of the reaction of SOR and SOD with superoxide, respectively. The cte term represents the rate of synthesis of $O_2^-$ by the xanthine oxidase system. Under these conditions, when the initial rate of oxidation of Center II is decreased by 50% due to the competition with SOD for $O_2^-$, it can be written[2] :

$$k_{SOD}[SOD] = k_{SOR}[SOR] \quad (5)$$

The concentration of SOD which decreases by 50% the rate of oxidation of Center II is graphically determined from Eq. (4), as illustrated in Figure 3. Taking into account the known second order rate constant of the reaction of $O_2^-$ with SOD ($2\times10^9$ $M^{-1}$ $s^{-1}$ for the CuZn-SOD from bovine erythrocytes or $3.2\ 10^8$ $M^{-1}$ $s^{-1}$ for Fe-SOD from *E. coli*) the second order rate constant of the oxidation of Center II by $O_2^-$ can be now calculated using Eq. (5). A value of $7\times10^8$ $M^{-1}$ $s^{-1}$ was obtained for the SOR from *D. baarsii*.

With the same assay, the SOR from *T. pallidum* exhibits a slight higher value of the rate constant for its reaction with $O_2^-$ of $1\times10^9$ $M^{-1}$ $s^{-1}$.[4]

# Conclusions

So far there is no suitable enzymatic assay for following purification of a SOR enzyme from cell extracts. However, the protein can be purified in all the cases according to its specific absorbance band at around 650 nm when oxidized why a strong oxidant (ferricyanide, for example).

The overall SOR activity (Eq.(3)) is probably rate limited by the efficiency of the electron donors ((Eq. (2)). However, taking into account the great instability of $O_2^-$ in solution (spontaneous dismutation rate constant of $5\times10^5$ $M^{-1}$ $s^{-1}$) and that dehydratase enzymes like aconitase and fumarase react very rapidly with superoxide (rate constant of about $10^7$ $M^{-1}$ $s^{-1}$)[11], the efficiency of SOR as an antioxidant resides in the very large rate constant for its reaction with $O_2^-$ (Eq. 1). A slower reaction rate for Eq.(1) would probably make an useless SOR enzyme, which would have no chance to react with $O_2^-$ within the cell.

# References

[1] I. Fridovich, *Ann. Rev. Biochem. 64*, 97 (1995).

[2] M. Lombard, D. Touati, Fontecave M., and V. Nivière, *J. Biol. Chem. 275*, 27021 (2000).

[3] F. E. Jr. Jenney, M. J. M. Verhagen, X. Cui, and M. W. W. Adams, *Science 286*, 306 (1999).

[4] M. Lombard, M. Fontecave, D. Touati, and V. Nivière, *J. Biol.Chem. 275*, 115 (2000).

[5] T. Jovanovic, C. Ascenso, K. R. O. Hazlett, R. Sikkink, C. Krebs, R. Litwiller, L. M. Benson, I. Moura, J. J. G. Moura, J. D. Radolf, B. H. Huynh, S. Naylor, and F. Rusnak, *J. Biol. Chem. 275*, 28439 (2000).

[6] J. K. Voordouw, and G. Voordouw, *Applied Env. Microb. 64*, 2882 (1998).

[7] M. J. Pianzzola, M. Soubes, and D. Touati, *J. Bacteriol. 178*, 6736 (1996).

[8] P. Tavares, N. Ravi, J. J. G. Moura, J. LeGall, Y. H. Huang, B. R. Crouse, M. K. Johnson, B. H. Huynh, and I. Moura, *J. Biol. Chem. 269*, 10504 (1994).

[9] A. V. Coelho, P. Matias, V. Fülöp, A. Thompson, A. Gonzalez, and M. A. Coronado, *J. Biol. Inorg. Chem. 2*, 680 (1997).

[10] P. Y. Andrew, Y. Hu, F. E. Jenney, M. W. W. Adams, and D. C. Rees, *Biochemistry 39*, 2499 (2000).

[11] D. H. Flint, J. F. Tuminello, and M. H. Emptage, *J. Biol. Chem. 268*, 22369 (1993).